\documentclass[journal abbreviation, manuscript] {copernicus}

\newcommand\hlinewidth[1]{\noalign{\hrule height #1}}
\newcommand{\customfigure}[3]{\begin{figure}%
\centering
\includegraphics[scale=#1]{#2.pdf}%
\centering%
\caption{#3}\label{fig:#2}%
\end{figure}}

\begin{document}


\title{Can repeller dynamics explain dominant pebble axis ratios?}

\Author[1,2]{Balázs}{Havasi-Tóth}

\affil[1]{Budapest University of Technology and Economics, Department of Fluid Mechanics, Bertalan Lajos u. 4-6. H-1117 Budapest, HUNGARY}
\affil[2]{HUN-REN-BME Morphodynamics Research Group, Budapest University of Technology and Economics,  Műegyetem rkp. 3. H-1117 Budapest, HUNGARY}
\correspondence{Balázs Havasi-Tóth (havasi-toth.balazs@gpk.bme.hu)}

\runningtitle{TEXT}
\runningauthor{TEXT}

\received{}
\pubdiscuss{} 
\revised{}
\accepted{}
\published{}

\firstpage{1}

\maketitle
\begin{abstract}
In the present work we focus on the morphology of well abraded, almost perfectly ellipsoidal natural pebbles. Flat, oblate and prolate ellipsoidal pebbles are expected to be characterized by qualitatively different shape evolution due to being unequally prone to specific rolling and sliding motions in natural environments. The three typical motions are considered as sliding on the flattest side and rolling around the longest and shortest axes. First, we assume by mechanical considerations that each of these motions alone affects the shortening of only one or two principal axes, then we formulate a mathematical expression associating the willingness as a probability to each motion of a given ellipsoid with arbitrary axis ratios. As a consequence, any of the motions is capable of suppressing the other two when the pebble is either ideally flat, oblate or prolate.

We present through a numerical analysis that our model predicts a self-excited shape evolution and curiously suggests the existence of an unstable equilibrium (repeller) in the vicinity of the naturally dominant average axis ratios found in the literature. Due to the self-excited evolution, pebbles diverging away from the proximity of the repeller are expected to rapidly become sufficiently thin or flat to easily break up into smaller fragments, which are no longer considered relevant in our investigation. Ultimately, our model outlines the possibility that the dominant axis ratios might be due to a well-balanced mixture of the motions and the corresponding slow evolution around the repeller.

Besides the theoretical predictions, we present experimental investigations of the dominant motions of various 3D-printed ellipsoidal shapes over an inclined planar surface and compare the results with the predictions of our model. We found that the observed behaviors not only reproduced the expected dominant motions but the non-trivial transition between the two relevant types of rolling motions has been captured as well.
\end{abstract}


\introduction \label{sec:intro}  

In the lack of a universal and widely accepted explanation of the naturally dominant axis ratios, the shape of rounded (all sharp edges and corners completely worn out) fluvial and coastal pebbles attract the interest of researchers even today. One of the most remarkable experimental sampling of coastal pebbles was carried out by \citet{Carr1969}, who investigated the shapes of almost 100.000 pebbles collected along the Chesil Beach, Dorset, United Kingdom. Similarly to \citet{Landon1930}, he arrived at the conclusion, that an \textit{equilibrium shape} retained by the pebbles during the abrasion should exist. Later, the topic was frequently visited through a sequence of short notations initiated by \citet{Wald1990}. Based on carefully selected set of well worn symmetrical pebbles, he claimed that the principal axis ratios appear to settle around the \textit{ultimate stable} proportion 7:6:3. However, \citet{Ashcroft1990} pointed out and \citet{Yazawa1990} agreed that anisotropies in the composition of the pebbles might have a decisive effect on the limit shape. The spatial segregation of the pebbles over a beach by shape and size was discussed by \citet{Lorang1990}. They pointed out that flattened pebbles at sandy beaches tend to travel -- by sliding on the flattest side -- shorter distances per wave periods than the spherical ones, leading to the accumulation of flat pebbles at the shoreline, while the frequency of the spherical ones is higher in deeper regions.

\customfigure{0.6}{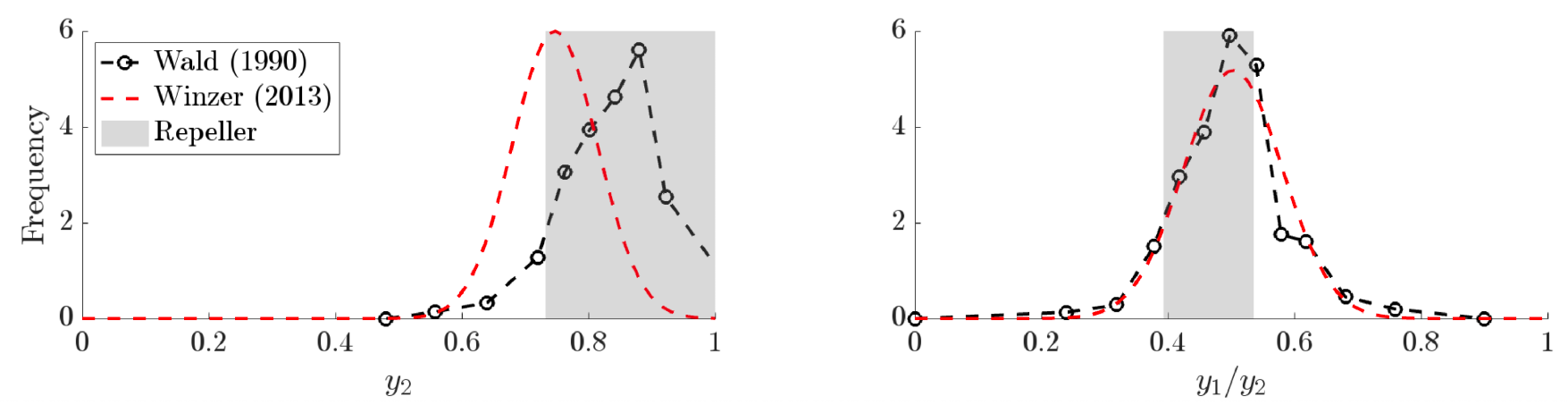}{Comparison of the normalized probability density functions of naturally abraded almost perfectly ellipsoidal pebbles \citep{Winzer2013,Wald1990}. The shaded regions mark the possible locations of the repeller introduced in the current paper.}

Including similar observations in terms of size- and shape-dependent motions, the classification of the most significant pebble behaviors was formed by \citet{Winzer2013}, who also made theoretical investigations referred later in this section. Samplings made by them on the beaches of the Cape Verde Islands (Sal and Boa Vista), the Canary Island of La Palma and along the Turkish coast near Alanya found slightly different ratios compared with Wald's 7:6:3 proportion. 
The probability density function (PDF) of the pebble axis ratios found by Winzer and Wald is shown in Figure \ref{fig:exp_compare} by the interpretation of the principal axis lengths as $u_1<u_2<u_3$ of the pebbles. The axis ratios then can be written as
\begin{equation}
\begin{split}
y_1 = \frac{u_1}{u_3}, \quad y_2 = \frac{u_2}{u_3}, \quad y_3 = u_3.
\end{split}
\end{equation}
The repeller marked with the shaded areas is of great significance in the present work and will be discussed in Section \ref{sec:lab_experiments}.

The theoretical descriptions of the evolution of coastal and fluvial pebble shapes are known to be under discussion since Aristotle \citep{Krynine1960,Domokos2012}. Assuming that the pebble shapes are the results of abrasion and complex material transport, Aristotle proposed a geometrical approach to the pebble shape evolution based on the geometrical properties of individual pebbles. He assumed the abrasion speed  of a surface point of a convex pebble to be the function of its distance $r$ from the pebble's center of gravity. Today, his concept could be expressed as
\begin{equation} \label{eq:aristotle}
\frac{\partial r}{\partial t} = f(r),
\end{equation}
which is the so-called \textit{distance-driven} flow of abrasion, where $f(r)$ is a monotonically increasing function of $r$ and $\frac{\partial r}{\partial t}$ is parallel with the surface normal.

The modern era of the investigation of natural pebble shapes has been commenced by the experimental and theoretical work of Lord Rayleigh in the 1940's \citep{Rayleigh1942,Rayleigh1944}. Based on his measurements, he observed that the limit shape produced by abrasion is not necessarily spherical. He proved that the sufficient condition for a pebble to evolve homothetically in terms of its principal axis ratios is the rate of ablation being  the $1/4$th power of the Gauss curvature $K=\kappa\lambda$:
\begin{equation}
    v=K^{1/4},
\end{equation}
where $\kappa$ and $\lambda$ are the principal curvatures. As a part of his attempts to find attrition processes that produce spherical shapes, he constructed experiments in a vertical cylindrical water-filled vessel (simulating a natural pothole) and a co-axially rotating paddle. The vortex induced by the paddle carried marble specimens round the vessel over the abrasive planar bottom. The internal wall of the vessel was initially smooth. He observed that no progress has been made in the vessel towards the sphere even in case of almost spherical initial shapes. Oblate shapes became more oblate, prolate ones became more prolate, which experimentally suggests that a self-excitation process can occur during the abrasion.

Assuming that the Gauss curvature is the major reason of the abrasion, Firey proved in 1974 that the limit shape of a convex symmetrical pebble under collisions with very large incoming particles from uniformly distributed directions is spherical \citep{Firey1974}. Firey's model can be expressed as
\begin{equation} \label{eq:Firey}
v=cK,
\end{equation}
where $c$ is a constant. Later, the generalization of Firey's model to non-spherical shapes was carried out by \citet{Andrews1999}. The extension of the abrading particles to arbitrary sizes was carried out by \citet{Bloore1977}, who proposed the equation
\begin{equation} \label{eq:Bloore}
v=1+2bH+cK,
\end{equation}
where $b$ and $c$ are constants, $H$ and $K$ are the mean and Gauss curvatures respectively.

Since the observations indicate that the occurrence of spherical pebbles is very rare in the nature, numerous attempts were made to capture the key effects explaining the dominant pebble axis ratios. Apparently, the amount of mathematical concepts of collisional abrasion is persistently increasing, while the effects of sliding and rolling friction are much less discovered. It has been suspected that the frictional processes deviate the pebble evolution from the spherical shapes, hence more recent investigations are focusing on the discovery of frictional effects \citep{Domokos2012,Winzer2013,Winzer2021,Hill2022}. Winzer observed that coastal pebbles on inclined sandy beaches are susceptible to show a stable rolling motion in shallow water flows of relatively high kinetic energy suggesting a non-spherical limit shape. Most recently, Hill proposed a contact likelihood function that quantifies the probability of ablation over the surface of a pebble as the function of its shape by implying the curvature as well as the excitation wave series and minimal potential energy need for a pebble to roll. His results show considerable agreement with Rayleigh's experiments.

In attempt to simplify Bloore's partial differential equation (PDE) in (\ref{eq:Bloore}), \citet{Domokos2012} deduced the curvature-driven box equations by assuming the evolution of ellipsoidal pebbles and implying collisional and frictional abrasion as well. The box equations have the same structure for collisions as (\ref{eq:Bloore}) but operate on the curvatures of ellipsoids corresponding to the principal directions. Being a system of ordinary differential equations (ODE's), the complexity of the solution is significantly reduced. After showing that the box equations reproduce the behavior of (\ref{eq:Bloore}) in terms of collisional abrasion, Domokos and Gibbons constructed a generalization of Aristotle's distance-driven model to imply frictional effects and managed to find critical points at arbitrary axis ratios. In Table \ref{tbl:models}, we present how the most significant abrasion models found in the literature predict the evolution of pebble axis ratios in the plane of axis ratios also known as Zingg's triangle \citep{Zingg1935}. In this paper we introduce our mathematical model shown in the last column of the table.

\renewcommand{\arraystretch}{1}
\begin{table}[h!]
\caption{Trajectories of the abrasion models in Zingg's triangle.}\label{tbl:models}
  \centering
  \begin{tabular}{ | m{2.5cm} | m{2.5cm} | m{2.5cm} | m{2.5cm} | m{2.5cm} | m{2.5cm} |}
    \hline
    &
    \begin{minipage}{.3\textwidth}
        \vspace{5pt}
        \includegraphics[width=25mm, height=25mm]{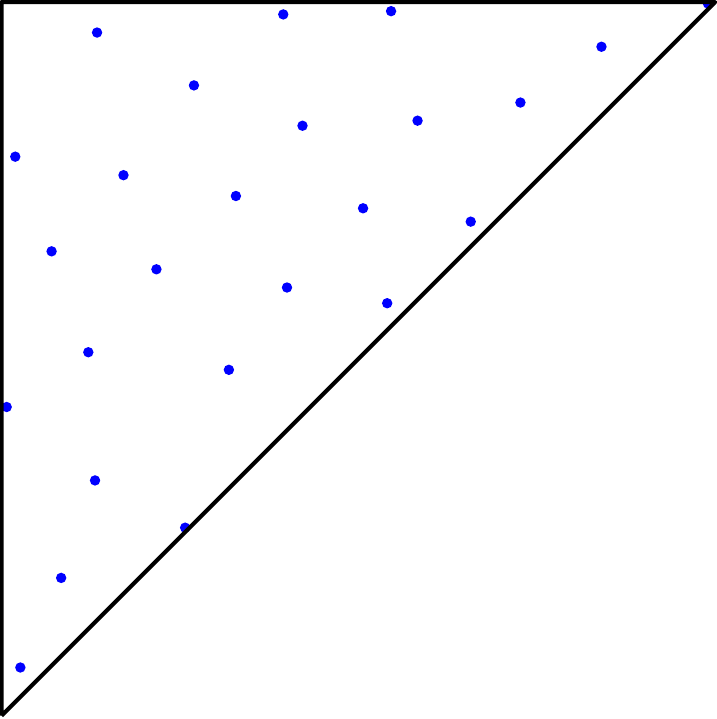}
        \vspace{5pt}
    \end{minipage}&
    \begin{minipage}{.3\textwidth}
        \vspace{5pt}
        \includegraphics[width=25mm, height=25mm]{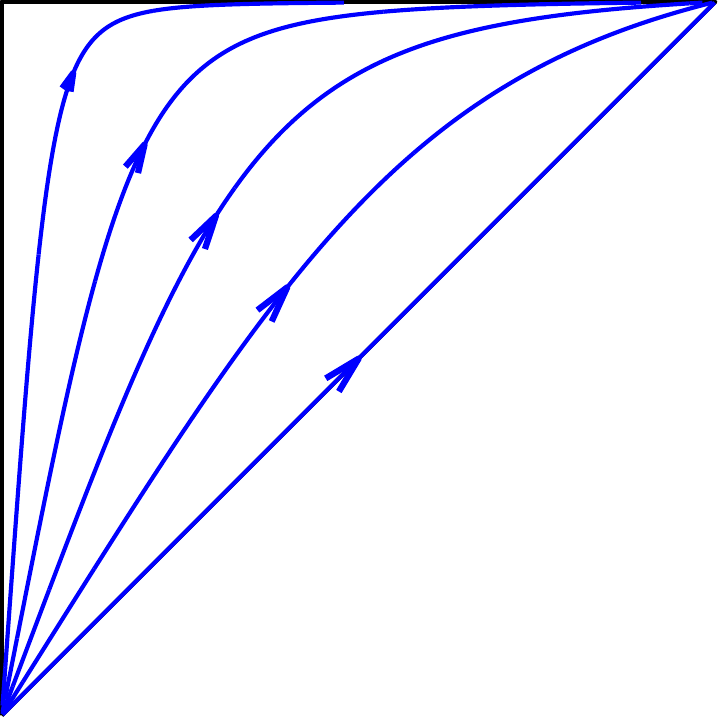}
        \vspace{5pt}
    \end{minipage}&
    \begin{minipage}{.3\textwidth}
        \vspace{5pt}
        \includegraphics[width=25mm, height=25mm]{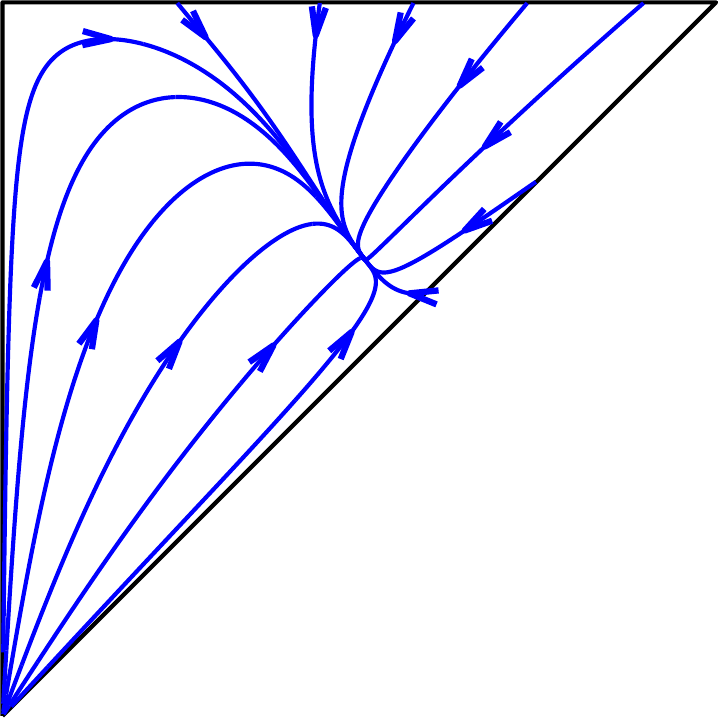}
        \vspace{5pt}
    \end{minipage}&
    \begin{minipage}{.3\textwidth}
        \vspace{5pt}
        \includegraphics[width=25mm, height=25mm]{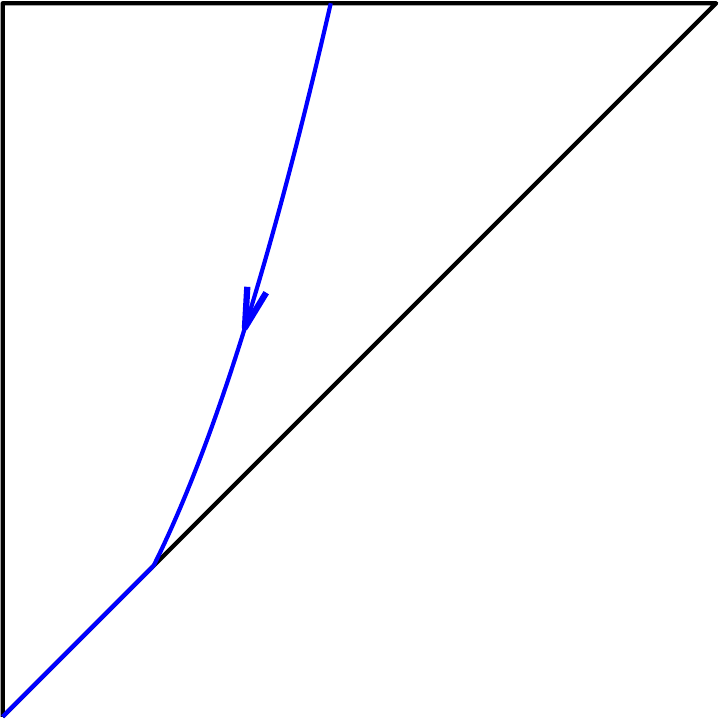}
        \vspace{5pt}
    \end{minipage}&
    \begin{minipage}{.3\textwidth}
        \vspace{5pt}
        \includegraphics[width=25mm, height=25mm]{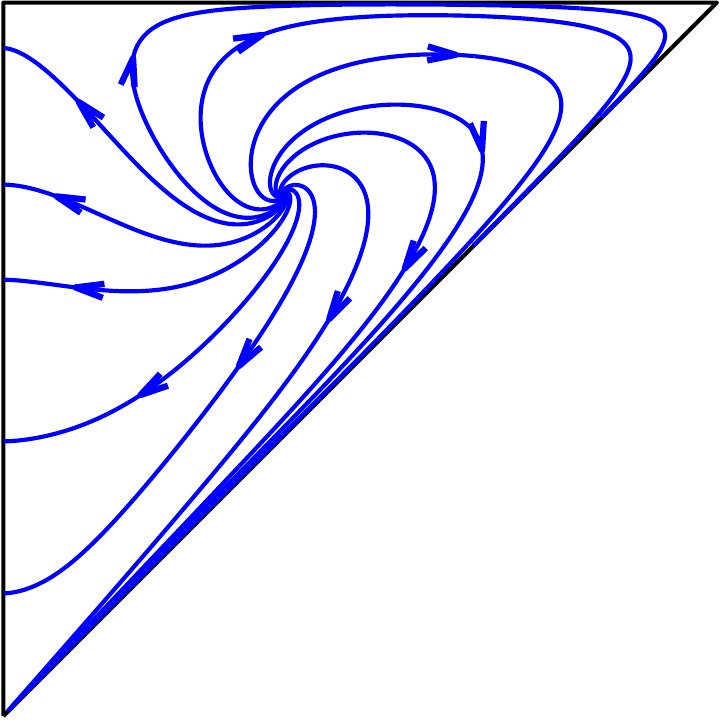}
        \vspace{5pt}
    \end{minipage} \\
    \hline
    Limit shape & All homothetic & Sphere & Arbitrary & Prolate & Flat and prolate\\
    \hline
    \hline
    \citet{Krynine1960} (Aristotle) & $f=ar$ & $f(0)=f'(0)=0$ & $f=\alpha r+\beta r^2$ & & \\
    \hline
    \citet{Rayleigh1942} & $v=K^{1/4}$ & & & & \\
    \hline
    \citet{Firey1974} & & $v=cK$ & & & \\
    \hline
    \citet{Domokos2012} & & $\nu_R=0, \nu_S=0$ & $\nu_R>0, \nu_S>0$ & & \\
    \hline
    \citet{Winzer2021} & & & & $y_2=4.73y_1^2$ & \\
    \hline
    Self-excited abrasion model & & & & & $\nu_S^k+\nu_R^k+\nu_W^k=1$ \\
    \hline
  \end{tabular}
\end{table}

The widely accepted conviction on the existence of a stable equilibrium shape induced by the observations in the nature appears to be an obvious and rarely questioned deduction. Moreover, it can be noted that theoretical approaches also adopt this assumption and -- motivated by the relatively sharp distributions (cf. Figure \ref{fig:exp_compare})-- make attempts to verify the conclusions by finding stable limit shapes using a great variety of mathematically or physically motivated abrasion models. In the present paper, this generic concept will be queried by the proposal of our alternative frictional abrasion model, that suggests an evolution of the opposite direction. We aimed to improve the frictional box equations through the introduction of an intuitive theory of curvature-driven frictional abrasion that is fundamentally different from the former explanations of the dominance of non-spherical pebbles. We show how the pebble shapes might be evolved through a self-excitation phenomenon leading to the presence of an unstable equilibrium point in Zingg's triangle. Instead of the investigation of the direct modeling of the pebble motion as Winzer and Hill captured the possible reasons of non-uniform ablation, we focus on the typical axis ratio dependent behavior of pebbles.

The paper is organized as follows: in Section \ref{sec:novel_approach} and \ref{sec:methodology} we introduce our concept and theoretical model, which encounter the box equations with our proposal of axis ratio-dependent pebble motions. The analytical and numerical investigation of the governing equation can be found in Section \ref{sec:numerical_results}. In Section \ref{sec:lab_experiments} we present our experimental layout and measurement results on capturing the non-trivial transition between different rolling motions. Finally, the discussion of the introduced model is closed with the conclusion in Section \ref{sec:conclusion}

\section{Frictional abrasion associated to typical pebble motions} \label{sec:novel_approach}
Recent studies started investigating on the effects of sedimentary rock (\citet{Cassel2021}) and grain (\cite{Deal2023}) shapes in flow induced transport. \citet{Cassel2021} showed that the mobility of well rounded symmetrical pebbles significantly depends on the axis ratios of the pebbles. Formerly, \citet{Winzer2013} observed and pointed out that the three main types of ellipsoidal coastal pebble motions under periodic wave excitation can be distinguished as:
\begin{itemize}
    \item Type I: pebbles remain stationary or slide a few centimeters during each excitation period,
    \item Type II: rolling motion around the longest axis,
    \item Type III: rolling motion around the shortest axis, dominated by inertial forces.
\end{itemize}
Although Winzer's classification focusing on the excitation energy levels are useful to better understand the abrasion processes on shingle and sandy beaches, the influence of the particular pebble shapes on the compatibility with the motions was not investigated.

As a possible geometrical explanation of Winzer's observations, we suggest the association of the pebble shapes at the three edges of Zingg's triangle with the listed motion types. Illustratively, Figure \ref{fig:boundary_flow} shows how typically flat blade (the vertical edge), prolate (diagonal edge) and oblate (horizontal edge) pebbles, might prefer the three motion types respectively. More precisely, flat blade, prolate and oblate shapes are more likely to show sliding motion on the flattest side (Type I), rolling around the longest (Type II) or shortest axis (Type III). In contrast with the motion types I-II, the Type III motion -- described in detail by \citet{Winzer2013} -- is a more complex sliding and rotating mixture of motions dominated by inertial forces and governs the shortening of the two largest axes of an oblate pebble. It should be noted here, that the actual axis shortening theory proposed by \citet{Winzer2013} was later shown to be incorrect by \citet{Winzer2021}. Nevertheless, it does not reduce the value of their observations and the resulting classification.

\customfigure{0.4}{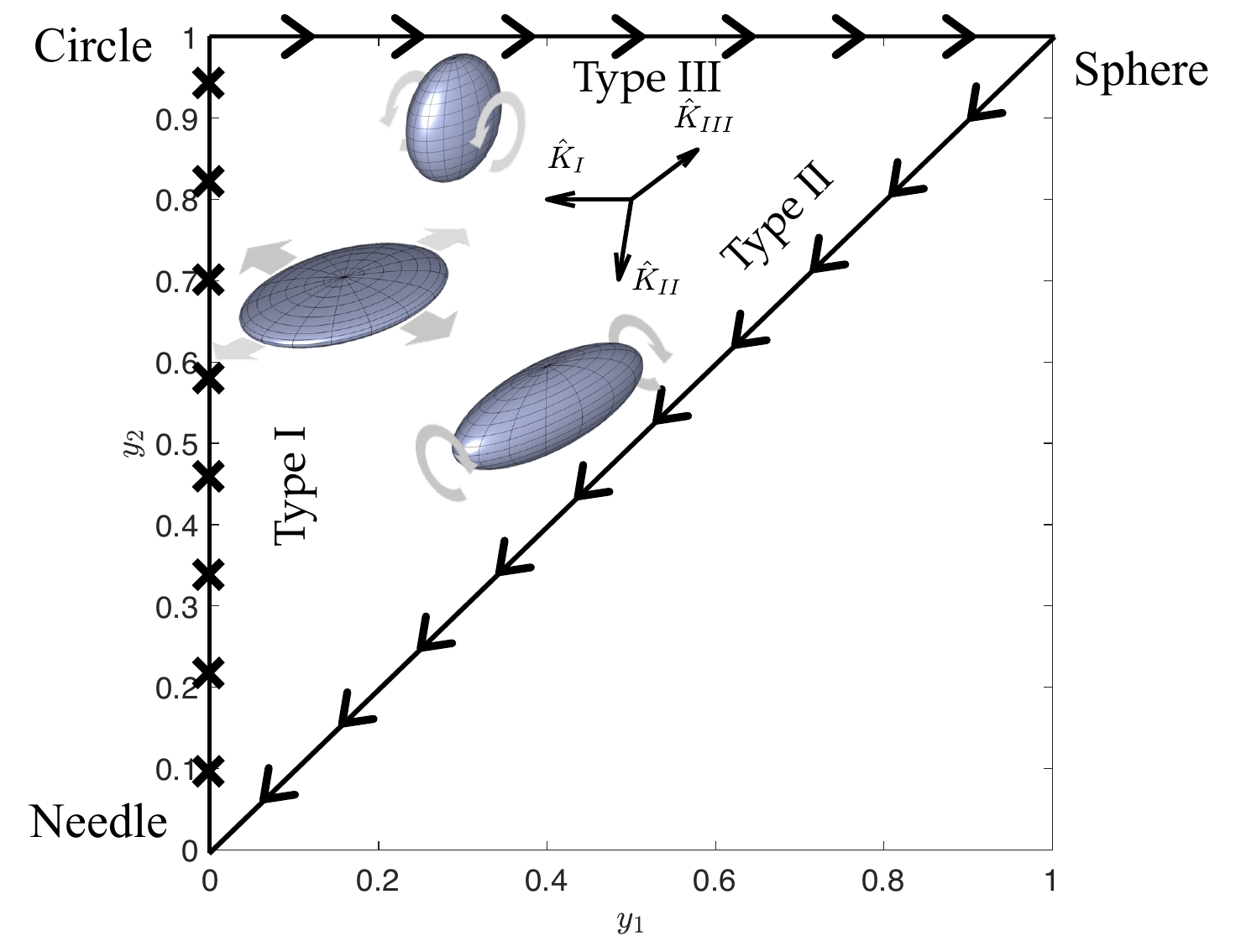}{Evolution expected in the case of perfectly flat, oblate, and prolate shapes around the edges of Zingg's triangle. Arrowheads and 'x' marks show the flow direction and stationary states respectively.}

The listed motion types and the implication of the aforementioned shape-dependency lead to exciting consequences. Firstly, each of the motions can be considered accountable for dominantly shortening either a single or a pair of principal axes through sliding or rolling frictional abrasion. Furthermore, similarly to one of Rayleigh's findings \citep{Rayleigh1944}, the motion types are expected to evolve the pebble shapes so that they become more and more prone to be subjects of the dominating motion. The Type I sliding motion shortens only the smallest axis ($u_1$), hence flat pebbles are expected to become more and more stable for sliding. Similarly, Type II shortens the two smaller axes ($u_1$ and $u_2$) through a rolling motion that leads to a more prolate shape. Due to the shortening of the two longer axes ($u_2$ and $u_3$), Type III predicts a less straightforward abrasion process. Based on mechanical considerations, \citet{Winzer2013} noted that as long as $u_1<<u_2\approx u_3$, the largest principal moment of inertia -- parallel with $u_1$ -- dominates the dynamics of the rolling motion. However, it is foreseeable, that the continuation of the rolling motion gradually diminishes its dominance as $u_2\rightarrow u_1$, until the motion becomes untenable, and another motion type must take place. A possible transition process can be outlined by means of the evolution according to the corresponding axis shortening process. For this purpose, we plotted the expected flow of flat, oblate and prolate ellipsoids on the edges of Zingg's triangle in Figure \ref{fig:boundary_flow}. The diagram suggests a back-turning flow at the top right corner, which phenomenon will be discussed later in Section \ref{sec:numerical_results}. Together with the expected transition, the described positive feedback could play a decisive role during the shape evolution and motivated the construction of our selective -- in terms of the principal axes shortening -- curvature-driven frictional abrasion model.

\section{The mathematical model} \label{sec:methodology}
The curvature-driven collisional box equations proposed by \citet{Domokos2012} provide qualitatively correct solutions to abrasion of well rounded ellipsoidal shapes using a remarkably simpler system of ODE's compared with the PDE introduced by \citet{Bloore1977}. Here we adopt the same geometrical concept in terms of the curvature-driven shortening of the principal axes of ellipsoids; however, we imply the selective axis shortening with respect to the aforementioned motion types. Summarizing the previous section, we introduce our mathematical model focusing on the fulfillment of the following considerations:
\begin{itemize}
    \item flat blade, prolate and oblate pebbles are specialized to carry out motion types I-III respectively,
    \item each motion type shortens only the one or two principal axes being in contact with the river bed or coastal substrate,
    \item the motions of pebbles occupying the edges of Zingg's triangle are exclusive.
\end{itemize}

Firstly, we aim to determine how a pebble with given axis ratios is prone to show each of the motion types. In accordance with Figure \ref{fig:boundary_flow}, we define three bounded linear kernel functions to the motion types for the sake of simplicity as
\begin{equation}\label{eq:nu_functions}
\begin{split}
& \text{Type I: }\nu_{S}(u_1,u_3)=1-\frac{u_1}{u_3}=1-y_1\quad \text{(=1 if the pebble is ideally flat),}\\
& \text{Type II: }\nu_{R}(u_1,u_2,u_3)=1+\frac{u_1}{u_3}-\frac{u_2}{u_3}=1+y_1-y_2\quad \text{(=1 if the pebble is prolate),}\\
& \text{Type III: }\nu_{W}(u_2,u_3)=\frac{u_2}{u_3}=y_2 \quad \text{(=1 if the pebble is oblate)},
\end{split}
\end{equation}
so that they individually estimate the probabilities of different motions in the function of the axis ratios. The subscripts $S, R$ and $W$ refer to the three motion types in order. Assuring exclusivity, we assume that ideally flat ($y_1=0$), prolate ($y_1=y_2$) and oblate ($y_2=1$) pebble shapes only prefer the single corresponding motion. Thus we define a power $k$ so that
\begin{equation} \label{eq:k_eq}
\nu_{S}^k+\nu_{R}^k+\nu_{W}^k=1,
\end{equation}
which allows only one of the terms to become unity at the edges of the triangle by oppressing the other two. Subsequently, the terms on the left hand side of (\ref{eq:k_eq}) will be used as probabilities of the selective shortening. Attempting the construction of a mathematical model that qualitatively outlines the trends in the evolution, the unit of the axis length measures are chosen to be one (i.e., non-dimensional).

After obtaining the probabilities of the motions, we conceptualize the frictional box equations of motion-dependent abrasion. We consider the shape evolution in Zingg's triangle driven by the superposition of the weighted axes shortening effects so that
\begin{equation}\label{eq:box_eqs_full_ratios}
-\dot y_i=\nu_{S}^k\hat K_{\mathrm{I}i}+\nu_{R}^k\hat K_{\mathrm{II}i}+\nu_{W}^k\hat K_{\mathrm{III}i}, \quad (i=1,2,3)
\end{equation}
where $\dot y_i$ is the temporal derivative of $y_i$, $K_N\ (N=\mathrm{I},\mathrm{II},\mathrm{III})$ are the vectors consisting of the shape-relevant Gauss curvatures at the principal axes and $\hat K$ denotes vector normalization. Together with the probabilities in (\ref{eq:k_eq}), the vectors $K_N$ are responsible for the selective shortening.

Since (\ref{eq:box_eqs_full_ratios}) is written in the frame of the axis ratios, we compute the curvatures $K_N$ according to the transformation shown by \citet{Domokos2012}. The curvatures at the three principal axes are expressed as
\begin{equation}
    G_1=\frac{u_1^2}{u_2^2u_3^2}=\frac{y_1^2}{y_2^2y_3^2}, \quad G_2=\frac{u_2^2}{u_1^2u_3^2}=\frac{y_2^2}{y_1^2y_3^2},\quad G_3=\frac{u_3^2}{u_1^2u_2^2}=\frac{1}{y_1^2y_2^2y_3^2}.
\end{equation}
Then, after transforming to the frame of axis ratios we arrive at
\begin{equation}
    K_{\mathrm{I}}=\frac{1}{y_3}\begin{bmatrix} G_1 \\ 0 \\ 0 \end{bmatrix}, \quad K_{\mathrm{II}}=\frac{1}{y_3}\begin{bmatrix} G_1 \\ G_2 \\ 0 \end{bmatrix},\quad K_{\mathrm{III}}=\frac{1}{y_3}\begin{bmatrix} 0 \\ G_2 \\ y_3G_3\end{bmatrix}-\begin{bmatrix} \frac{1}{y_1y_2^2y_3^3} \\ \frac{1}{y_1^2y_2y_3^3} \\ 0\end{bmatrix}.
\end{equation}
The directions of the $\hat K_N$ vectors at a typical internal point is shown in Figure \ref{fig:boundary_flow}.
Note that the coefficients $1/y_3$ and the second term of $K_{\mathrm{III}}$ are present due to the conversions
\begin{equation}
\begin{split}
    &\dot u_1=\dot y_1y_3+\dot y_3y_1,\\
    &\dot u_2=\dot y_2y_3+\dot y_3y_2,\\
    &\dot u_3=\dot y_3.
\end{split} 
\end{equation}
Although the system (\ref{eq:box_eqs_full_ratios}) provides three dimensional trajectories in the $[y_1,y_2,y_3]$ space, we hereinafter omit the shrinking of the pebbles by setting the constraint $\dot y_3=0$. Thus, we focus on a restricted planar flow in the $[y_1,y_2]$ plane, which is however, dependent on the pebble size $y_3$. Due to the preservation of the longest axis, we refer to our model with this assumption as a \textit{conservative} evolution, which significantly simplifies the dynamics without losing the key characteristics of the evolution. On the other hand, the evolution after setting the constraints $\dot y_1=\dot y_2=0, \dot y_3\neq 0$ in (\ref{eq:box_eqs_full_ratios}) would become homothetic in the function of $y_1, y_2$ and $y_3$.

\subsection{Stationary solutions and numerical results} \label{sec:numerical_results}
Due to the lack of closed solutions of (\ref{eq:k_eq}), the ODE system (\ref{eq:box_eqs_full_ratios}) cannot be investigated analytically in a generic case. Furthermore, the solutions do not exist at the vertices of Zingg's triangle. However, the three edges can be tested analytically for possible equilibrium points (where $\dot y_1=\dot y_2=0$) due to the elimination of two of the $\nu_N$ functions in the limit according to (\ref{eq:k_eq}). Since the curvature vectors $K_N$ are normalized in (\ref{eq:box_eqs_full_ratios}), the magnitude of the three terms on the right hand side are determined by the $\nu_N^k$ coefficients. With this in mind, we found that $y_1=0$ is unconditionally stable due to $K_I=0$ and $\nu_S=1$, whereas $y_2=1$ becomes a stable equilibrium if $y_3\rightarrow \infty$. We summarize the search for equilibrium points along the edges of Zingg's triangle in Table \ref{tab:equilibria}. 

Our findings suggest the $y_1=0$ line to be stationary at $0<y_2<1$ due to the vanishing curvature at the shortest axis of completely flat shapes. Although Table \ref{tab:equilibria} suggests no further stationary points along the triangle's hypotenuse, it can be shown that none of the valid solution trajectories leave Zingg's triangle and the flow at the boundary matches the expectations illustrated in Figure \ref{fig:boundary_flow}. Apart from the boundaries, we test the flow numerically in the followings with $\nu_N^k\neq 1$.
\begin{table}
\caption{\label{tab:equilibria} Search for equilibrium points along the edges of Zingg's triangle}
\begin{tabular}{ c c c c c c c}
\hlinewidth{1pt}
  & Constraint & $\nu_S^k$ & $\nu_R^k$ & $\nu_W^k$ & Restriction of (\ref{eq:box_eqs_full_ratios}) & Equilibrium points\\ 
  \hline
 Flat & $y_1=0$ & 1 & 0 & 0 & $\hat K_I=0$ & $y_1=0$ \\  
 Prolate & $y_1=y_2$ & 0 & 1 & 0 & $\hat K_{\mathrm{II}}\neq 0$ & none \\ 
 Oblate & $y_2=1$ & 0 & 0 & 1 & $\hat K_{\mathrm{III}} = [0,0,1]^T$ & $y_2=1$ if $y_3\rightarrow \infty$ \\
 \hlinewidth{1pt}
\end{tabular}
\end{table}

During the numerical investigation, we applied the variable order Adams-Bashforth-Moulton method with adaptive step size to solve (\ref{eq:box_eqs_full_ratios}) and approximated the solution of (\ref{eq:k_eq}) for $k$ using the Newton-Raphson method at every evaluation step. The detailed flow governed by (\ref{eq:box_eqs_full_ratios}) is shown in Figure \ref{fig:vflow}. Colored curves represent the isocurves of the velocity magnitude $\vert\dot y_i\vert$ and gray trajectories trace the evolution with different initial conditions. The set of the latter illustrates the flow inside the triangle, for which a single unstable spiral fix point (repeller) was found and marked with the black circles, while the $y_1=0$ axis is an attractor as expected for all trajectories that do not touch any of the edges of the triangle. The latter case is an extreme situation where $k\rightarrow\infty$, hence the flow of (\ref{eq:box_eqs_full_ratios}) simplifies to one of the analytically examined restricted systems (Table \ref{tab:equilibria}). Changing the pebble size $y_3$ shifts the fix point in the triangle as shown in Figure \ref{fig:vflow}b without affecting the structure of the flow; its infimum was found approximately at [$y_1=0.392, y_2=0.732$] as $y_3\rightarrow 0$, which is the case in Figure \ref{fig:vflow}a.

\customfigure{0.45}{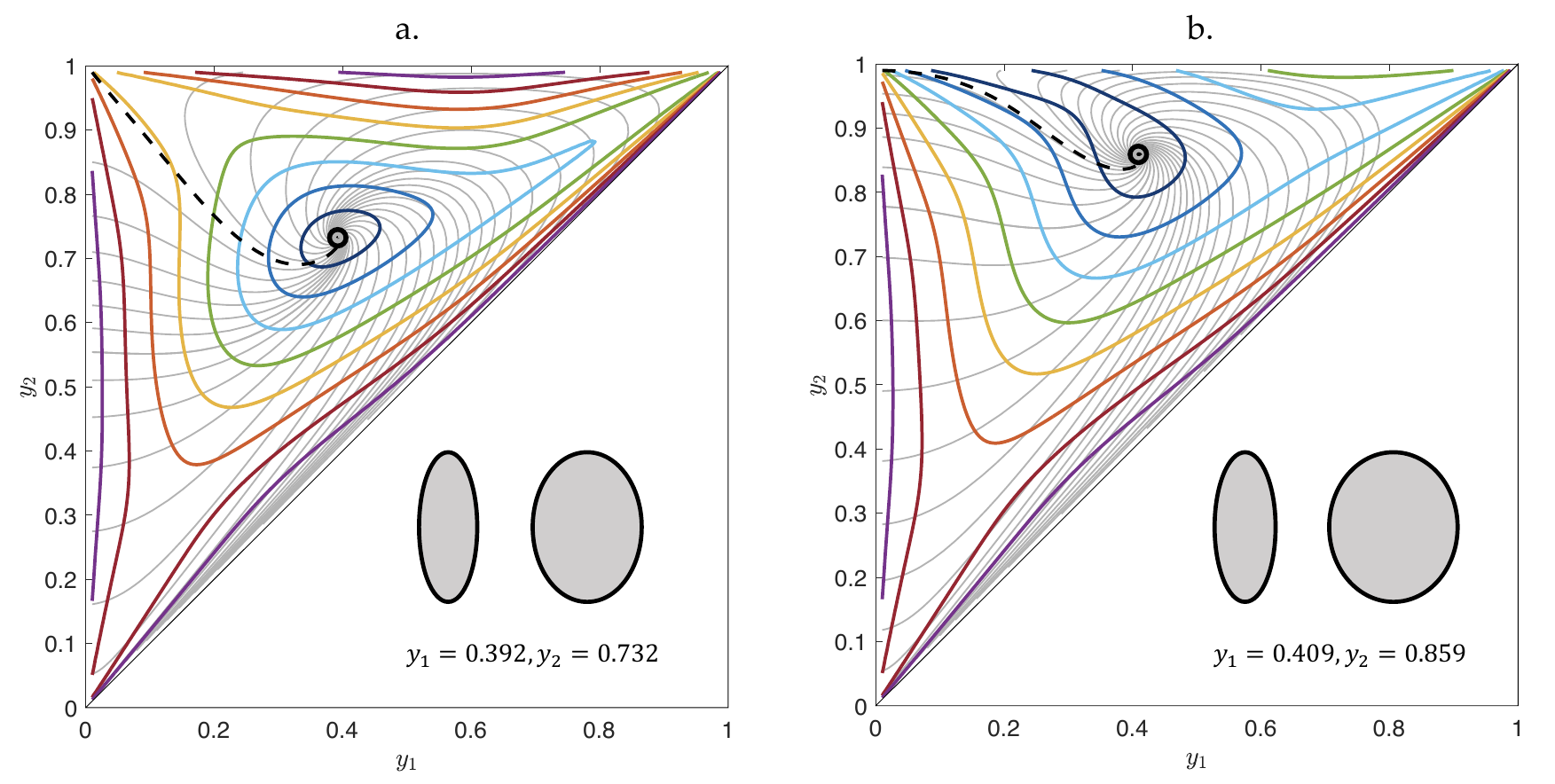}{Flow over Zingg's triangle in case of two different pebble sizes. a: $y_3\rightarrow 0$, b: $y_3=1.1$. Colors show the velocity magnitude ($\vert\dot y_i\vert$) isocurves increasing from the repeller (blue and purple curves represent slow and fast evolution respectively), gray curves show the trajectories of (\ref{eq:box_eqs_full_ratios}). The black circle marks the corresponding repeller. Finally, the dashed line is a separatrix interconnecting the saddle at $[y_1=0, y_2=1]$ with the repeller. The ellipses at the white area illustrate the axis ratios of the pebbles at the repeller.}

\customfigure{0.45}{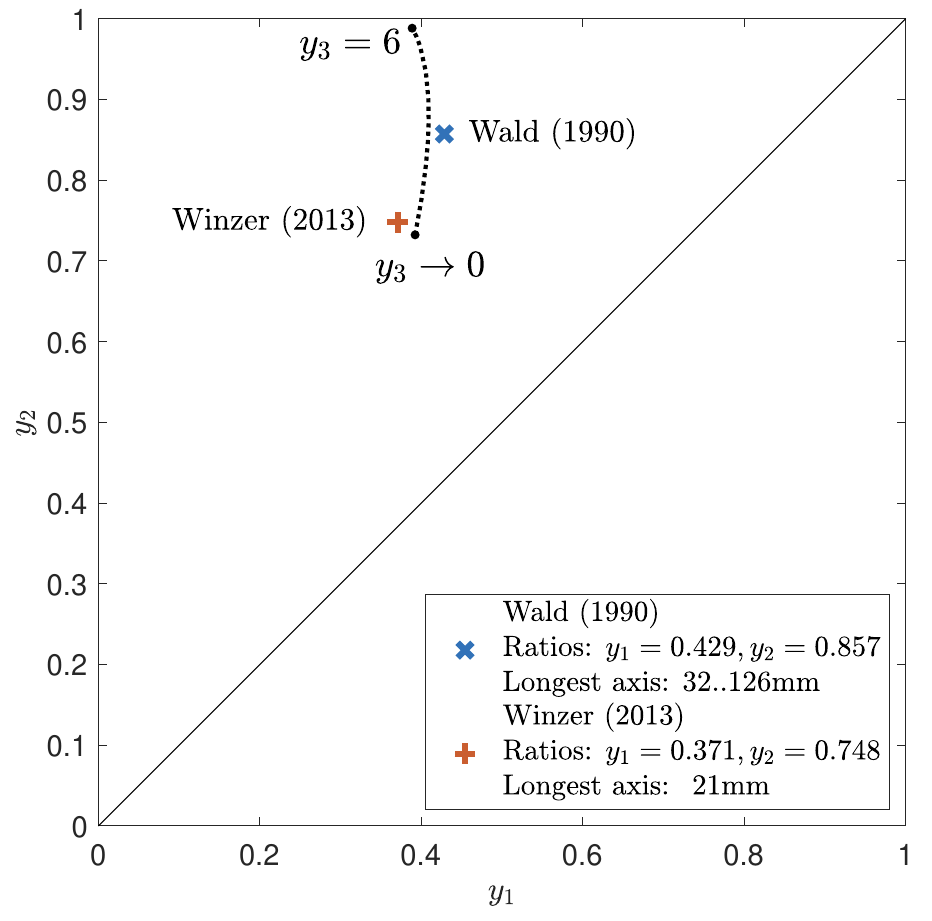}{Comparison of the possible locations of the unstable equilibrium point (dashed curve) in the function of the pebble size $y_3$ and the samplings of \citet{Wald1990} and \citet{Winzer2013} including significantly different pebble sizes. Again, the unit of measure of $y_3$ is considered unity.}

According to the presented sliding-rolling frictional abrasion model, the decisive ellipsoid axis ratios -- flat, prolate, and oblate -- become more and more dominant over time, leading to an interesting shape evolution of pebbles in the big picture. Due to the well-balanced behavior in terms of the motion types and abrasion, the evolution is rather slow in the vicinity of the repeller but rapidly accelerates while leaving its region as one of the motion types gradually becomes dominant. Consequently, the dynamics of the introduced model claims that the closer a pebble is to the repeller the longer it maintains its axis ratios through an almost homothetic evolution, while pebbles far enough away from the internal critical point become flatter or thinner and -- although it is not implied in the present model -- most probably breaks into smaller fragments.

We show a curve in Figure \ref{fig:fixpointcurve}, on which the repeller can exist as the pebble size changes. The curve runs between $y_3\rightarrow 0$ and $y_3=6$ considered as two reasonable limits. The location of the repellor in the limit $y_3\rightarrow \infty$ is hard to be computed numerically due to the diverging value of the power $k$ at the edge of the triangle. Nevertheless, it can be shown analytically that the evolution along the horizontal edge becomes homothetic when $y_3\rightarrow \infty$. In the same diagram, we placed the two significant results measured on symmetric and well abraded coastal pebbles sampled by \citet{Winzer2013} and \cite{Wald1990}. Apart from the geographical location, the main difference between the samplings was that Winzer focused on small pebbles of $21$ mm in average, while Wald's collection contained significantly larger specimens in the range of $32..126$ mm. It can be concluded, that not only both samplings lie remarkably close to the curve, but the difference in the pebble sizes show qualitative agreement as well. Note that due to the definition of the axis lengths, the $y_3$ values along the dashed line showing the set of possible locations of the repeller do not represent metric units of measure. The notably close match of the repeller and the samplings of natural pebble sets, together with the long residence time in the same region due to the slow abrasion rate $\dot y_i$ suggests a counter-intuitive explanation of the observations in the nature: although the equilibrium point is unstable, the collective set of abrading coastal pebbles might have the tendency to gather in the repeller's neighborhood.

Another remarkable peculiarity of the flow is the gradual transition from motion Type III to II in the vicinity of the top right corner. As it was discussed earlier in section \ref{sec:novel_approach}, the Type III motion at the top edge is accountable for the shortening of the two longest axes, leading to a rather spherical shape. Sooner or later, the intermediate axis should become equal to the shortest one. On the one hand, using our notation of the axis ratios $y_i$, the further shortening would require the switching of the shortest two axes, while on the other hand, Type III motion becomes less and less dominant due to the vanishing dominance of rotational stability around the shortest axis. Consequently, the aforementioned gradual transition traced by the back-turning trajectories inherently avoids such conditions and blurs the transition between the two rolling motion types. As an experimental verification of the existence of the back-turn, we aim to experimentally capture such a behavior in Section \ref{sec:lab_experiments}.

\section{Laboratory experiments} \label{sec:lab_experiments}
Throughout this section, we present the results of laboratory experiments carried out to capture relevant motions of rolling ellipsoids. One of our fundamental concepts in the present work was that the shortening of the principal axes due to friction is strictly driven by the shape-dependent motion types. Equivalently, we assume that the typical motion of a pebble determines the instantaneous direction of its evolution in Zingg's triangle. Therefore, instead of the direct measurement of the abrasion process, which is a significantly more complex and challenging task, here we only focused on the characteristic motions of ellipsoids with different axis ratios while the axis lengths remained intact.

\subsection{Experimental layout}
As an imitation of coastal and fluvial water-induced excitation of pebbles, we aimed to investigate the motion of various ellipsoids over an inclined plane with a predefined speed and angle of the slope. To simulate an infinite plane, we applied a conveyor belt shown in Figure \ref{fig:conveyor_belt_layout} and \ref{fig:conveyor_belt_images} with adjustable inclination and speed. The main parameters of the conveyor belt were as follows:
\begin{itemize}
    \item Inclination range: $0-30$°,
    \item Speed range: $0-1$ m/s,
    \item Length of the flat surface: $0.5$ m,
    \item Width of the flat surface: $0.15$ m.
\end{itemize}

\customfigure{0.3}{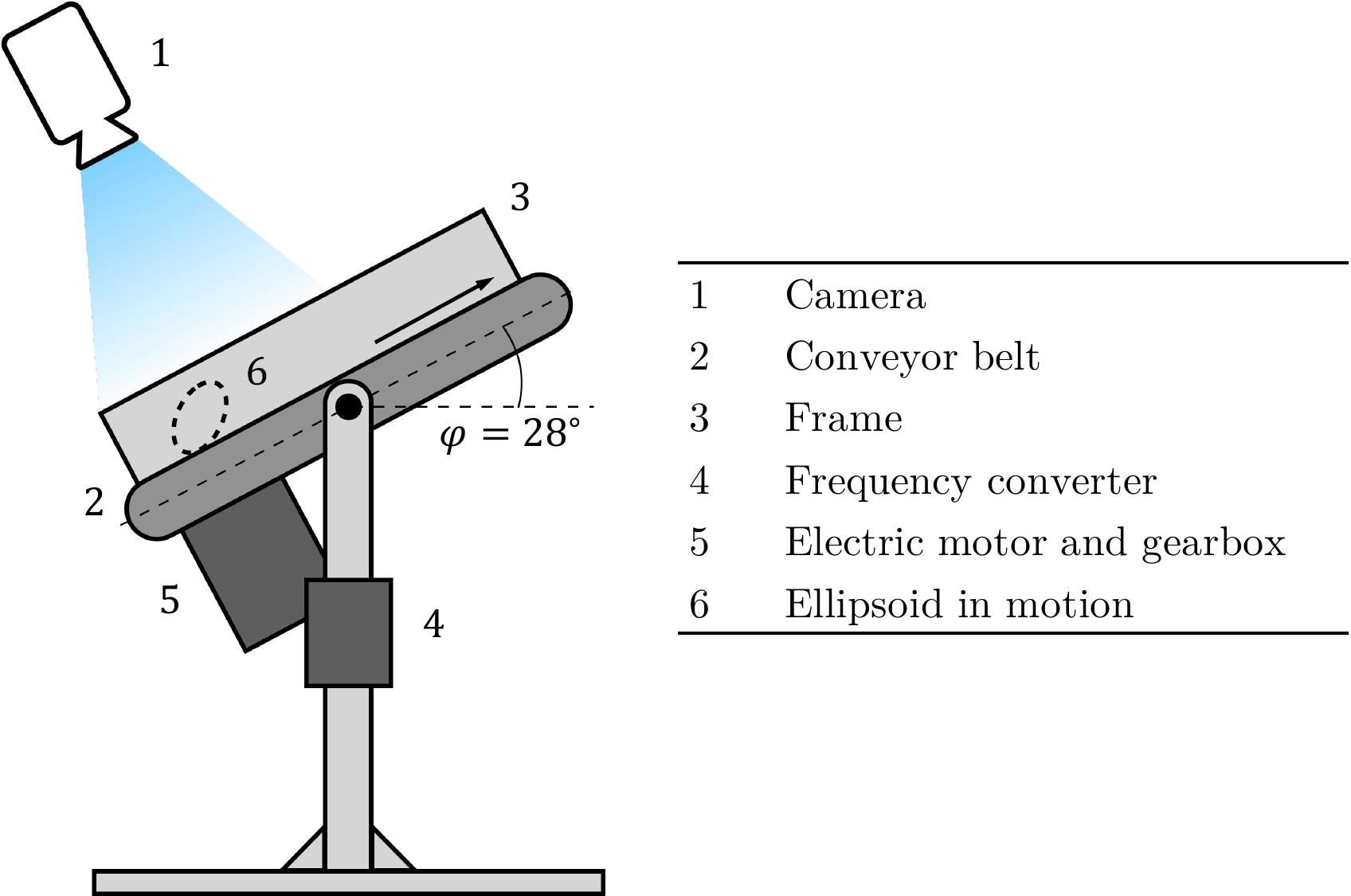}{Schematic layout of the experiments. The arrow shows the moving direction of the belt.}
\customfigure{0.5}{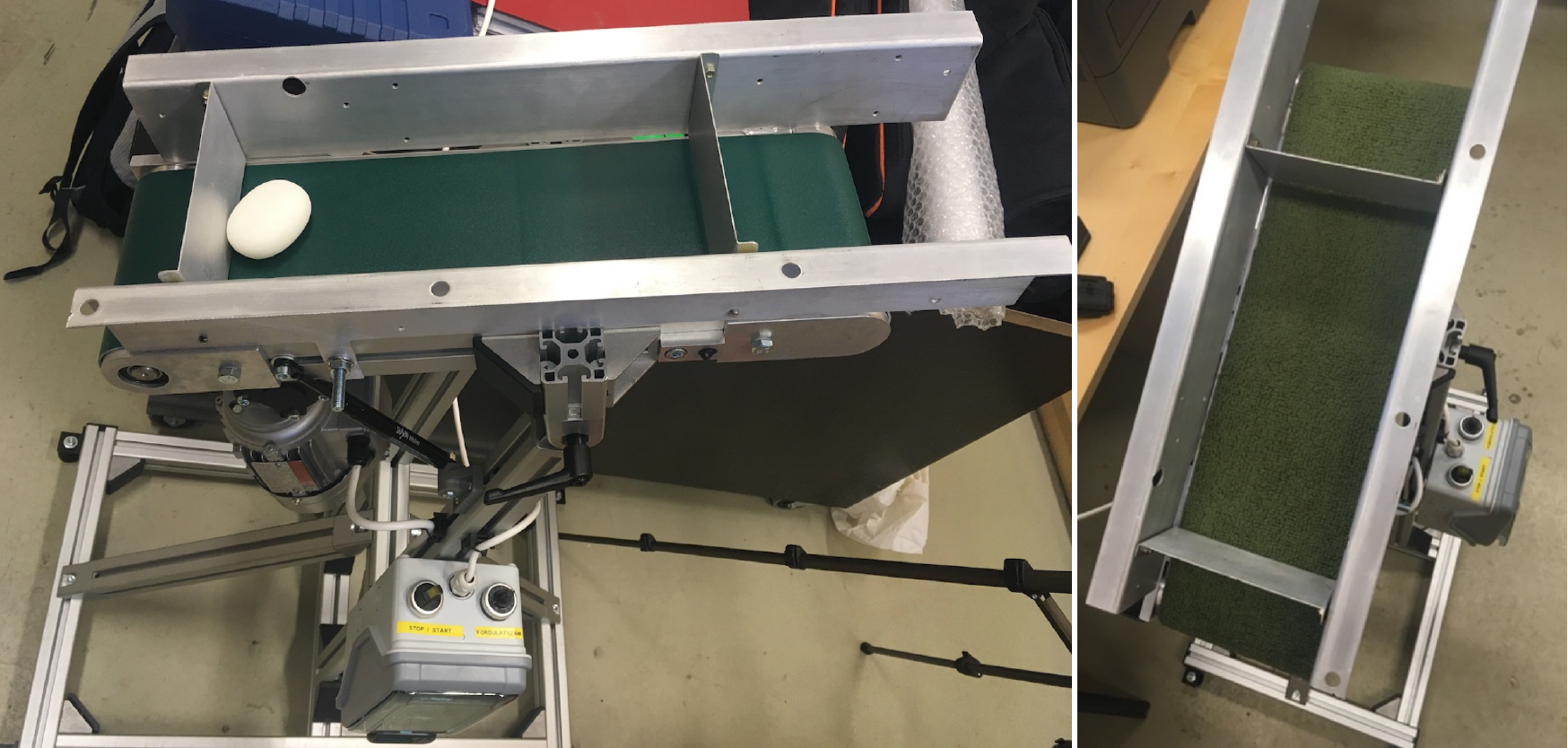}{The conveyor belt with adjustable inclination and running speed.}

The role of the frame built over the moving surface of the belt is to prevent the rolling ellipsoids from falling off the surface. Also, to damp the aleatory collisions with the belt and avoid the bouncing of the hard ellipsoids, the belt has been covered with a soft and moderately slippery material. During the operation of the conveyor belt, the motion of the investigated ellipsoid placed on its surface has been recorded using a 30 fps (frames per second) camera fixed $70$ cm above the surface of the belt.

\subsubsection{Selection of the ellipsoids and parameters}
To capture the discussed back-turn of the trajectories, we aimed to select a series of ellipsoidal shapes lying along a single trajectory of (\ref{eq:box_eqs_full_ratios}) with $y_3\rightarrow 0$. The motivation of this selection is that a series of evolutionary states of the same pebble would be tested. In order for a proper selection to be made, we formulated a simple approximate transition criterion based on the minimal excitation energy required for a pebble to roll around one of its axes. Since the rotation of a body around its second principal direction ($u_2$ axis in our case) is unstable, we simply omit the corresponding energy requirement, and only focused on the other two rotations (i.e., Type II and Type III motions).

In order for a pebble at rest on its flattest side to be able to roll around its longest axis, it has to be given a potential energy of at least
\begin{equation} \label{eq:energy_min3}
E_3=\frac{mg}{2}(u_2-u_1),
\end{equation}
where $m$ is the mass of the ellipsoid and $g$ is the gravitational acceleration. Similarly, the energy demand to roll around its shortest axis while being at rest on its intermediate axis can be expressed as
\begin{equation} \label{eq:energy_min1}
E_1=\frac{mg}{2}(u_3-u_2).
\end{equation}
Presumably, it is expected for the transition between the motion types III and II to take place where the two rolling motions equally likely occur, based on the energy demands. Thus, the energies \eqref{eq:energy_min3} and \eqref{eq:energy_min1} should be equal, which is satisfied when the three axis lengths form an arithmetic sequence so that
\begin{equation} \label{eq:ari_seq}
u_3-u_2=u_2-u_1,
\end{equation}
from which the critical line of the transition on the $[y_1,y_2]$ plane is expressed as
\begin{equation} \label{eq:transition_line}
y_2=\frac{1+y_1}{2},
\end{equation}
and shown in Figure \ref{fig:traj_ellipsoids_transition} with the dashed line.

\customfigure{0.5}{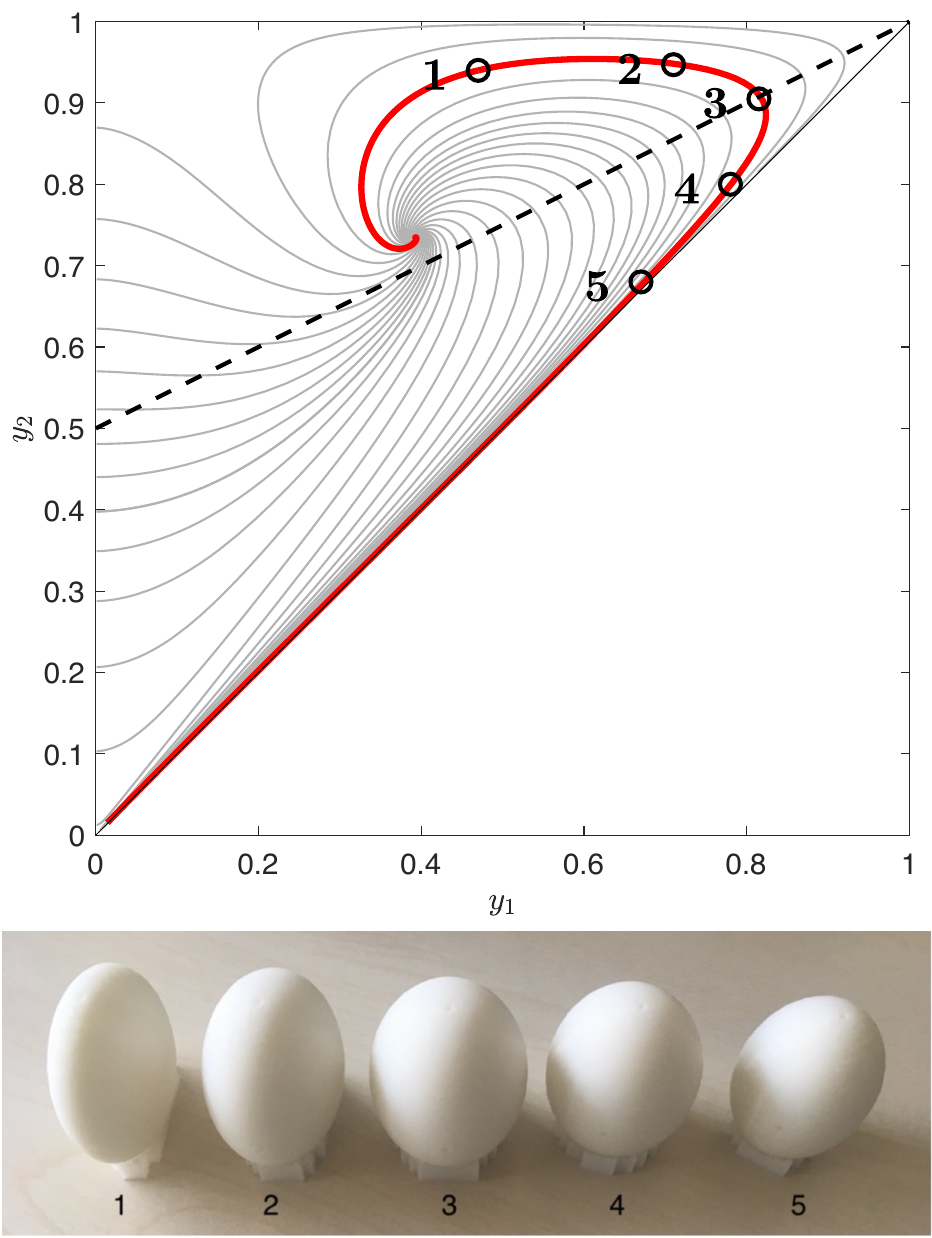}{Ellipsoids along the selected trajectory and the printed geometries associated with the selected axis ratios. The dashed line shows the estimated transition between motion types III and II.}

We chose five different ellipsoids along a selected trajectory in this region in such a way that the first and the last one settle far away from the aforementioned theoretical transition. The ellipsoids were prepared using an FDM-type 3D printer device with PLA filament and $0.2$ mm layer thickness. After printing, the surfaces of the ellipsoids were finished with fine sand paper to get rid of the anisotropic surface quality. The axis ratios of the investigated ellipsoids are listed in Table \ref{tab:axis_ratios}, while the lengths of the longest axes of the ellipsoids were set to $60$ mm.
\begin{table}
\caption{\label{tab:axis_ratios} The axis ratios of the investigated ellipsoids.}
\begin{tabular}{ c c c }
\hlinewidth{1pt}
$N$ & $\hspace{5mm} y_1$ & $\hspace{3mm} y_2$ \\
\hlinewidth{0.5pt}
1 & \hspace{5mm}0.47 & \hspace{3mm}0.94 \\
2 & \hspace{5mm}0.71 & \hspace{3mm}0.95 \\
3 & \hspace{5mm}0.82 & \hspace{3mm}0.91 \\
4 & \hspace{5mm}0.78 & \hspace{3mm}0.80 \\
5 & \hspace{5mm}0.67 & \hspace{3mm}0.68 \\
\hlinewidth{1pt}
\end{tabular}
\end{table}
For the sake of easier comparison and to emphasize the robustness of the evolution, we fixed the conveyor belt inclination and speed ($25$ degrees and $0.55$ m/s respectively) in case of each ellipsoid during the experiments. We assume that the identical parameters of the test equipment simulate the same environmental conditions and excitation energy levels for all ellipsoids of the same major axis size.

\subsection{Evaluation and discussion}
\customfigure{0.5}{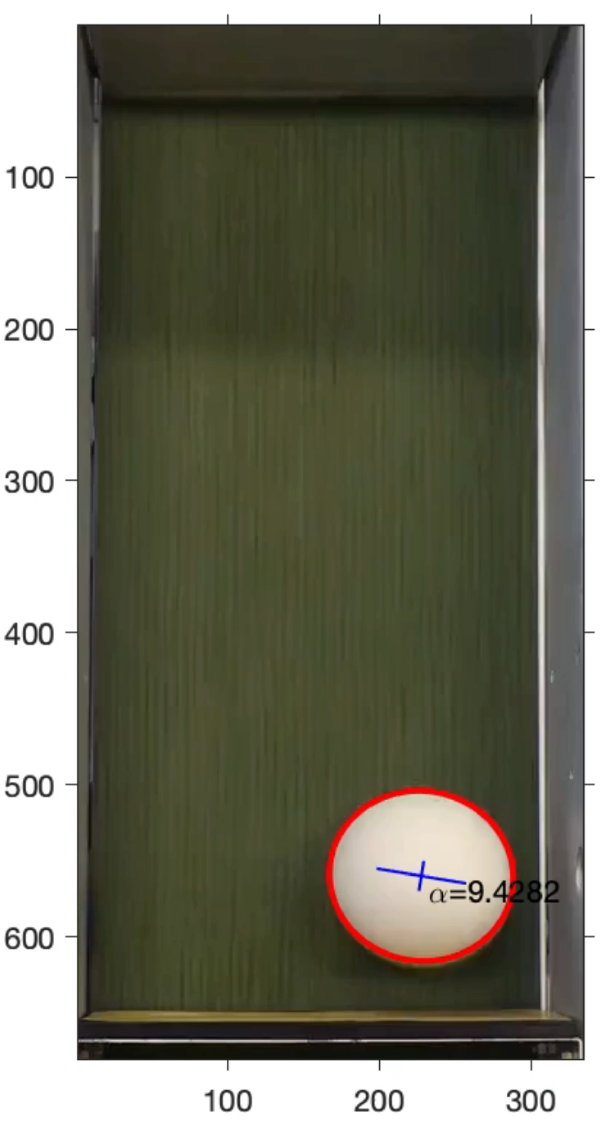}{Processing the instantaneous image of a moving ellipsoid over the surface of the belt. The red ellipse is fitted on the image of the ellipsoid, the blue lines show the directions of its principal axes. Axis labels indicate the pixel counts.}

\customfigure{0.67}{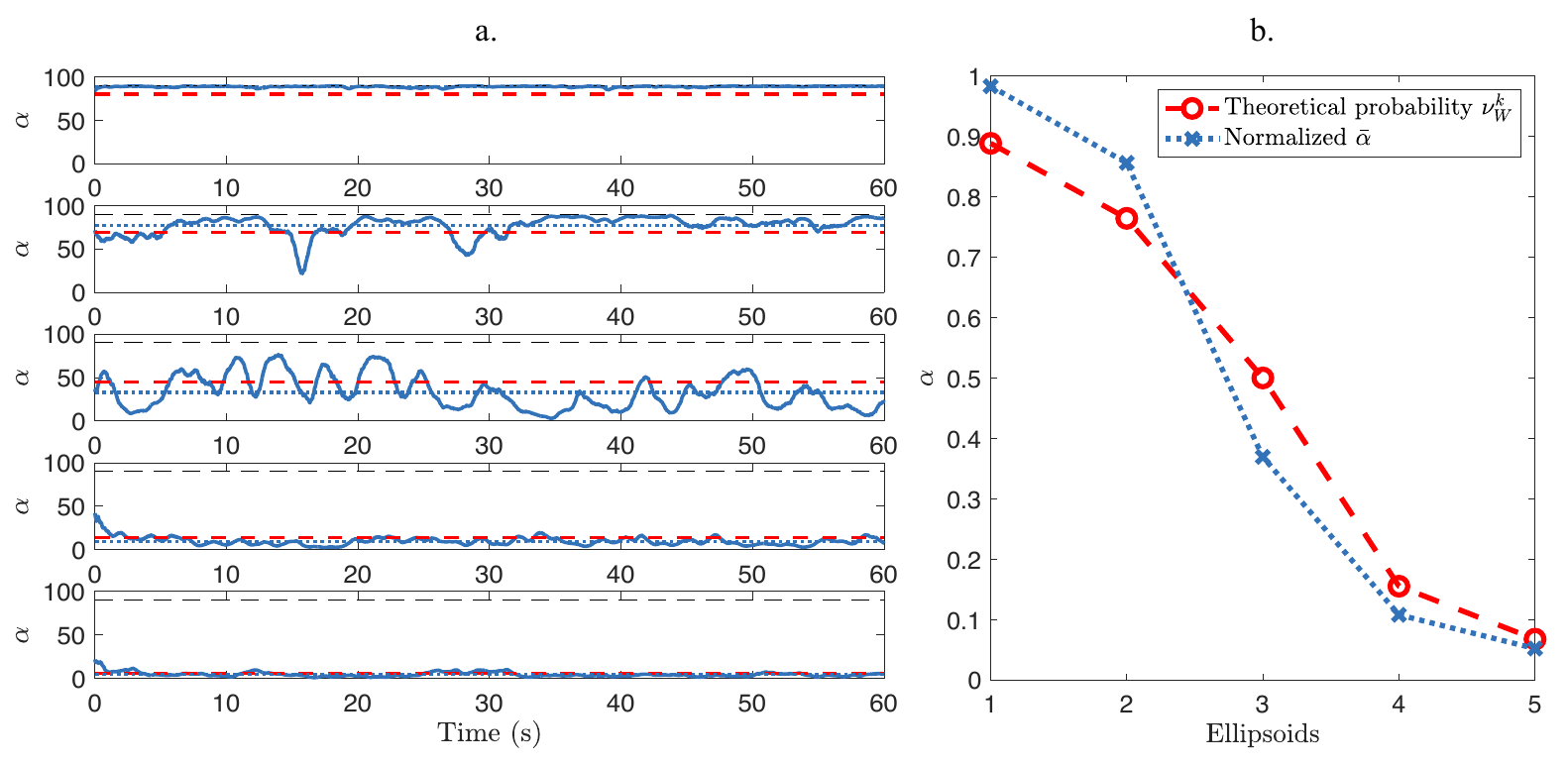}{a: Time series of the angle $\alpha$ of the major axes with blue solid line in case of the five ellipsoids (the first one is at the top). The thin black dashed line marks $90^{\circ}$, blue and red thick dashed lines show the experimental mean $\bar\alpha$ and theoretical probability $\nu_W^k$. b: Theoretical ($\nu_W^k$) and experimental probabilities of the rotation around the shortest axis.}

In order to avoid the interactions between the ellipsoids on the belt, the experiments were carried out independently. The motions of each of the five ellipsoids were recorded for a $60$ s time interval. Although the problem is three-dimensional, the pure motions of types I-III can be considered as planar. Consequently, types II and III can be distinguished without a complete three-dimensional reconstruction of the ellipsoids' motions. Therefore, the appropriate image processing consisted of an automated ellipse fitting on each of the instantaneous images. A typical image of the ellipse fitting with its computed principal axes directions is shown in Figure \ref{fig:image_proc}. Since the background was dark green, we identified the white ellipsoids simply by the pixels' brightness. Then the $\alpha$ angle between the horizontal direction and the major axis of the ellipse is computed at each frame, providing sufficient data to confidently distinguish motion types II and III by means of the expected values of $\alpha\approx 0°$ and $\alpha\approx 90°$ respectively.

\customfigure{0.5}{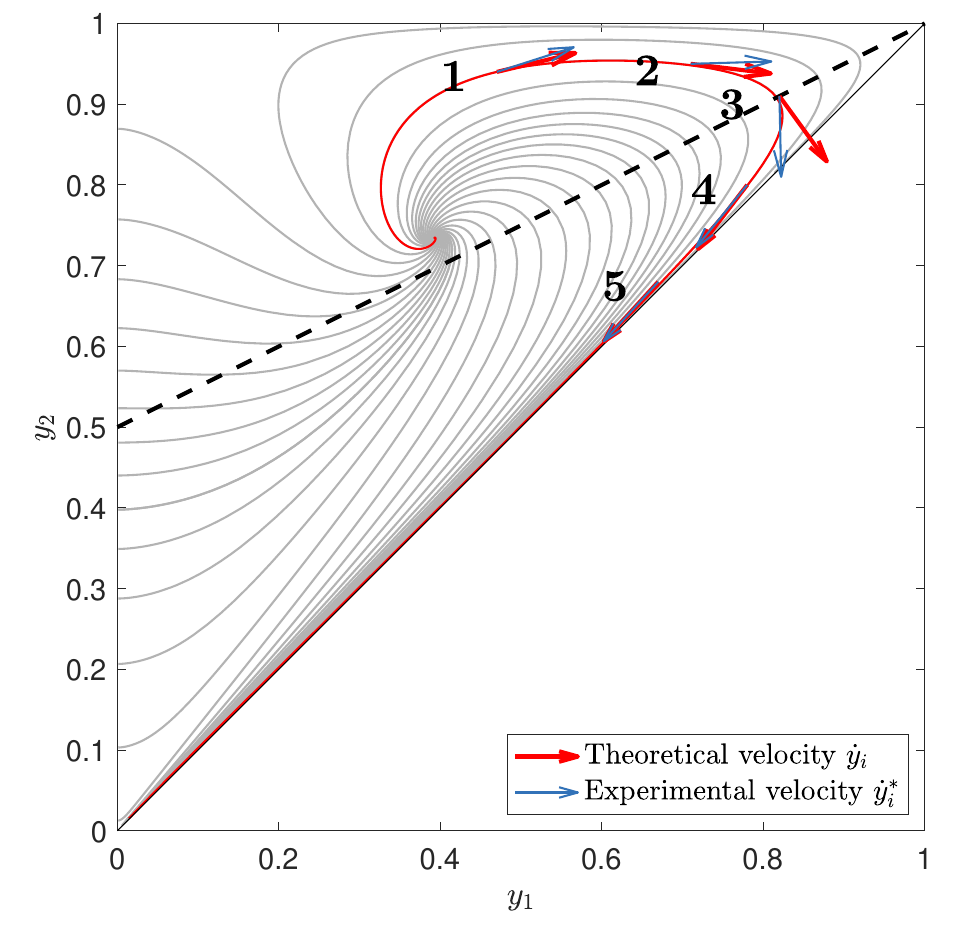}{Numerical (blue) and experimental (red) predictions of the ellipsoid evolution along the selected trajectory.}

The recorded time series of $\alpha$ are presented in Figure \ref{fig:time_series}a for the ellipsoids $1$ to $5$ in order. Apparently, the first and the fifth ellipsoids dominantly rotate around the shortest (Type III) and longest axes (Type II) respectively, which is in a very good agreement with the predicted motions (cf. Figure \ref{fig:boundary_flow} and \ref{fig:traj_ellipsoids_transition}). The other three ellipsoids from $2$ to $4$ show a gradual transition between the first and fifth time series. At the peak of the back-turn, the third ellipsoid is characterized by the most disordered motion among the five time series. The tendency shown by the motions provides a qualitative match with both the back-turn and the energy based estimate location of the transition (\ref{eq:transition_line}). Figure \ref{fig:time_series}b presents the theoretical $\nu_W^k$ and the normalized experimental probabilities $\bar\alpha$ of the rotation around the shortest axis.

Finally, as a rough estimation, we determine the instantaneous experimental velocity vectors $\dot y_i$ to identify the expected shape evolution of the five ellipsoids. As the basis of the comparison, the theoretical velocities are given by (\ref{eq:box_eqs_full_ratios}) and (\ref{eq:k_eq}). In the lack of observed sliding motions on the conveyor belt, we compute the experimental probabilities associated with the motion types II and III as
\begin{equation} \label{eq:experimental_probabilities}
\begin{split}
    &\nu_{S}^*=0, \\
    &\nu_{R}^*=1-\nu_W^*, \\
    &\nu_{W}^*=\frac{\bar\alpha}{90},
\end{split}
\end{equation}
respectively. Note that the blue dashed line shows $\nu_{W}^*$ in Figure \ref{fig:time_series}b. Since we considered the shape evolution to be governed by the motion probabilities, substituting (\ref{eq:experimental_probabilities}) into (\ref{eq:box_eqs_full_ratios}), the experimentally predicted local velocities $\dot y_i^*$ can be computed. In Figure \ref{fig:arrows}, we plotted the obtained normalized velocity vectors so that the directions of the expected evolution to be compared. A significant difference can only be observed at the peak of the back-turn, where the direction of the trajectory changes rapidly and a considerable qualitative match was found in terms of the existence of the back-turn.


\conclusions \label{sec:conclusion}  
In the present paper we proposed a physically motivated hypothetical approach on modeling the evolution of coastal and fluvial pebble shapes. Although several recent investigations (\citet{Domokos2012,Winzer2013,Winzer2021,Hill2022}) made attempts to explain the non-spherical dominant axis ratios of naturally abraded pebbles -- by anticipating the existence of an attractor --, a widely accepted conclusion has not been formulated yet.

Focusing on the effects of frictional sliding and rolling abrasion, we propose a mathematical model constructed according to the following assumptions:
\begin{itemize}
    \item flat, prolate, and oblate shapes most probably slide on the flattest side, roll around the longest and shortest axes respectively,
    \item the distinguished motion types only determine the shortening of the axes in contact with the bed,
    \item the axes shortenings are curvature-driven,
    \item any of the motion types can become exclusive when the pebble shape is perfectly flat, prolate or oblate.
\end{itemize}
Motivated by the box equations introduced by \citet{Domokos2012}, we constructed a set of ODE's that fulfills the listed assumptions and drives the shape evolution of well-abraded pebbles by determining the selective shortening of the principal axes. After an analytical and numerical investigation of the ODE's, we found that an unstable equilibrium point -- instead of a stable one -- exists near the axis ratios often observed in the nature. The existence of an unstable equilibrium with a slow, almost homothetic evolution in its neighborhood counter-intuitively suggests a long life-time of naturally dominant axis ratios, while pebbles leaving its proximity rapidly become flat or thin enough to potentially break. Moreover, the presented model shows qualitative agreement with the samplings made by \citet{Winzer2013} and \citet{Wald1990} not only in terms of the average axis ratios but concerning the pebble sizes as well.

Besides the analytical and numerical investigation of the governing equations, we constructed an experimental equipment and measurement layout to verify the expected motion types of ellipsoids with carefully selected axis ratios. We assumed that the typical motion of an ellipsoid determines the instantaneous direction of the evolution. Our findings show qualitative match with the expected behavior in the critical area, where the transition between the two types of rolling motion is expected to occur.

\begin{acknowledgements}
The initial conceptualization of the shape-dependent motions of well rounded pebbles with the possible existence of a self-excited evolution was originated by Gábor Domokos, who proposed the motion-dependent selective axis shortening and supported the construction of our mathematical model as well as the design process of the experimental layout with valuable personal consultations.\\
Project no. TKP-6-6/PALY-2021 has been implemented with the support provided by the Ministry of Culture and Innovation of Hungary from the National Research, Development and Innovation Fund, financed under the TKP2021-NVA funding scheme. This research was supported by the NKFIH Hungarian Research Fund Grant 134199 is kindly acknowledged.
The author expresses their appreciation to the Department of Polymer Engineering at the Budapest University of Technology and Economics for their contribution to our experimental investigations by providing us the desired 3D-printed ellipsoids.
\end{acknowledgements}

\competinginterests{}
The author declares that they have no concerning financial or non-financial competing interests.



\bibliographystyle{copernicus}
\bibliography{references.bib}

\end{document}